\documentclass[12pt, showpacs, aps, showkeys]{revtex4}
\usepackage{amsmath}
\usepackage{amssymb}
\begin{document}

\title{Area spectrum of near-extremal SdS black holes via the new interpretation of quasinormal modes}
\author{Wenbo Li}
\email{liwenbo10@yahoo.com.cn}
\author{Lixin Xu\footnote{Corresponding author}}
\email{lxxu@dlut.edu.cn}
\author{Jianbo Lu}
\email{lvjianbo819@163.com}

\affiliation{Institute of Theoretical Physics, School of Physics \&
Optoelectronic Technology, Dalian University of Technology, Dalian,
116024, P. R. China}

\begin{abstract}

Motivated by the recent work about a new physical interpretation of
quasinormal modes by Maggiore, we investigate the quantization of
near-extremal Schwarzschild-de Sitter black holes in the four
dimensional spacetime. Following Kunstatter's method, we derive the
area and entropy spectrum of near-extremal Schwarzschild-de Sitter
black holes which differs from Setare's result. Furthermore, we find
that the derived a universal area spectrum is $2\pi n$ which is
equally spaced.

\end{abstract}


\maketitle

\section{Introduction}

The quantization of the black hole horizon area is a fascinating
subject. Since an equally spaced entropy spectrum was firstly
predicted by Bekenstein in $1974$ \cite{beken1}, there have been
many attempts to derive the entropy spectrum directly from the
dynamical modes of the classical theory
\cite{beken2,louko,dolgov,barvin,kastrup,beken3}. However, there has
been little known about the direct physical connection between the
classical dynamical quantities that give rise to Bekenstein-Hawking
entropy and the corresponding microscopic degrees of the quantum
black hole. An important step in this direction was made by Hod
\cite{hod1} by a semiclassical consideration of the macroscopic
oscillation modes of black holes. In particular, he assumed an
equally discrete area spectrum and used the existence of a unique
quasinormal mode frequency in the large damping limit to uniquely
fix the spacing. Dreyer \cite{dreyer} demonstrated that Hod's result
could be recovered in loop quantum gravity if the relevant group is
taken to be $SO(3)$ rather than $SU(2)$. These developments have
spurred lots of subsequent activity
\cite{frit,kunst,motl,corichi,abdalla0,card1,card2,hod2,birmin,ploy,
setare1,lepe,setare2,hod3,keshet1,keshet2,daghigh}.

Recently, a new physical interpretation for the quasinormal modes of
black holes was presented by Maggiore \cite{maggiore}. According to
Maggiore's proposal, in order to overcome or at least alleviate some
difficulties raised by the Hod's proposal \cite{hod1} in the
interpretation of quasinormal frequencies, the Black hole
perturbations are modeled in terms of a collection of damped
harmonic degrees of freedom. In addition, he indicated that the real
frequencies of the equivalent damped harmonic oscillators were
$(\omega^2_R+\omega^2_I)^{1/2}$, rather than simply $\omega_R$.
Motivated by Maggiore's work, Vagenas \cite{vagenas} utilized the
new proposal to the interesting case of Kerr black hole and proposed
a new interpretation as a result of Maggiore's idea, for the
frequency that appears in the adiabatic invariant of a black hole.
In a more recent paper \cite{medved}, a universal form for the Kerr
and Schwarzschild quantum area spectra was established by Medved by
presenting a simple but vital modification to a recent treatment on
the Kerr (or rotating black hole) spectrum. Although the above
considerations are still somewhat speculative, they certainly
propose a reasonable physical interpretation on the spectrum of the
black hole quasinormal modes. It is ,therefore, of interest to study
the area spectrum of other black holes, in particular, near-extremal
black holes for which the quasinormal frequencies are quite
different from that of non-extremal black holes.

In this article our aim is to investigate the area and entropy
spectrum of near-extremal Schwarzschild-de Sitter black holes in
four dimensional spacetime by adopting Maggiore's proposal.
According to Maggiore's work, the frequency of the harmonic
oscillator is $\omega_0=(\omega^2_R+\omega^2_I)^{1/2}$. Although the
author of Ref. \cite{setare2} has discussed the area and entropy
spectrum, the most interesting case is that of highly excited
quasinormal modes, whose frequency is $\omega_0=|\omega_I|$ rather
than simply $\omega_0=\omega_R$. Since, for $\omega_I\gg\omega_R$,
this observation will change the physical understanding of the black
hole spectrum and examine the various results of the literature. In
the next section, we will try to derive the area and entropy
spectrum by extending Kunstatter's method.

\section{Quasinormal Modes of Near Extremal SdS Black Holes}

The near-extremal Schwarzschild-de Sitter (SdS) spacetime in four
dimensions is a non-trivial case with a non-asymptotically flat
spacetime. General SdS spacetimes have a metric of the form
\begin{eqnarray}
ds^2=-f(r)dt^2+f^{-1}(r)dr^2+r^2d\Omega^2_2,
\end{eqnarray}
with
\begin{eqnarray}
f(r)=1-\frac{2M}{r}-\frac{r^2}{L^2_{ds}},
\end{eqnarray}
where $M$ denotes the black hole mass and $L^2_{ds}$ is the de
Sitter curvature radius, which related to the cosmological constant
$\Lambda$ by $L^2_{ds}=3/\Lambda$. The spacetime possesses two
horizons: the usual black hole horizon locates at $r=r_+$ and the
cosmological horizon locates at $r=r_c$, where $r_+<r_c$. We assume
that the three roots of the equation $f(r)=0$ are $r_+$, $r_c$, and
$r_0$ respectively. In terms of these roots, $f(r)$ can be rewritten
as
\begin{eqnarray}
f(r)=\frac{1}{L^2_{ds}r}(r-r_+)(r_c-r)(r-r_0),
\end{eqnarray}
with $r_0=-(r_++r_c)$. In addition, $M$ and $L^2_{ds}$ as functions
of these roots can be expressed as
\begin{eqnarray}
L^2_{ds}=r_+^{2}+r_+r_c+r_c^{2},\nonumber\\
2ML^2_{ds}=r_+r_c(r_++r_c).\label{m}
\end{eqnarray}
Defined by the relation $\kappa_+\equiv\frac{1}{2}(df/dr)|_{r=r_+}$,
the surface gravity $\kappa_+$ can be written as
\begin{eqnarray}
\kappa=\frac{(r_c-r_+)(r_+-r_0)}{2L^2_{ds}r_+}.
\end{eqnarray}

Let us now specialize to a non-trivial case with a
non-asymptotically flat spacetime called the near-extremal SdS black
hole. As for this case, the cosmological horizon $r_c$ is very close
to the black hole horizon $r_+$. Hence, one can make the following
approximations:
\begin{eqnarray}
r_0\sim-2r_+,\;\;\;\;\;\;\;\;L^2_{ds}\sim3r^2_+,\;\;\;\;\;\;\;
\kappa_+\sim\frac{r_c-r_+}{2r^2_+}.\label{kappa}
\end{eqnarray}

Cardoso and Lemos studied firstly the analytical quasinormal mode
spectrum for the near-extremal SdS black hole \cite{card2}, and they
concluded that the asymptotic quasinormal frequencies of
near-extremal SdS black hole are given by the simple expression
\begin{eqnarray}
\omega=\kappa_+\bigg[\sqrt{\frac{\upsilon_0}{\kappa^2_+}-1/4}-i(n+1/2)\bigg],\;\;\;\;\;\;n=0,1,2,...\label{QNM1}
\end{eqnarray}
where
\begin{eqnarray}
\upsilon_0=\kappa^2_+l(l+1),
\end{eqnarray}
for scalar and electromagnetic perturbations, and
\begin{eqnarray}
\upsilon_0=\kappa^2_+(l+2)(l-1),
\end{eqnarray}
for gravitational perturbations, $l$ is the angular quantum number.
Very recently, by computing the Lyapunov exponent which is the
inverse of instability timescale associated with the geodesic
motion, Cardoso et al presented that quasinormal modes of black
holes are determined by the parameters of the circular null
geodesics in the eikonal limit ($l\gg1$) \cite{card3}. And then,
they found a simple analytical quasinormal modes for the
near-extremal SdS black holes in $d=4$:
\begin{eqnarray}
\omega_{BQNM}=\kappa_+[l-i(n+1/2)].\label{QNM2}
\end{eqnarray}

Though the form of Eq. (\ref{QNM2}) is similar with Eq.
(\ref{QNM1}), there exists an important difference, namely, Eq.
(\ref{QNM2}) is valid for the low-lying modes $n\ll l$ with $l\gg1$
while Eq. (\ref{QNM1}) is valid in the limit $n\rightarrow \infty$.
Here we are interested in highly excited black holes, hence the area
spectrum of near-extremal SdS black holes should be discussed in the
next section with the aid of Eq. (\ref{QNM1}) found in \cite{card2}.

\section{Area Spectrum of Near Extremal SdS Black Holes}

We now utilize a new physical interpretation of quasinormal modes
proposed by Maggiore \cite{maggiore} to the near-extremal SdS black
hole and attempt to derive the area spectrum by following
Kunstatter's method.

Given a system with energy $E$, one can show the first law of black
hole thermodynamics
\begin{eqnarray}
dM=\frac{1}{4}T_HdA,
\end{eqnarray}
here we adopt the mass definition $E=M$ according to the Ref.
\cite{abbott}. Employing Kunstatter's method \cite{kunst}, the
adiabatically invariant integral can be written as
\begin{eqnarray}
I=\int\frac{d E}{\omega(E)}.\label{action}
\end{eqnarray}
At this point we should clarify what the frequency $\omega$ in the
denominator should be. As for a harmonic oscillator, the frequency
is the vibrational frequency that corresponds to the system's energy
$E$ under a slow variation of a parameter related to the energy,
hence a small variation energy $dE$ can be created and the quantity
$E/\omega$ is an adiabatic invariant. According to Maggiore's
proposal, one has to model the Black hole perturbations in terms of
a collection of damped harmonic degrees of freedom if one wants to
avoid the difficulties raised by Hod's conjecture in the
interpretation of quasinormal frequencies \cite{maggiore}. In
addition, Maggiore indicates that the real frequencies of the
equivalent damped harmonic oscillators should be written as
\begin{eqnarray}
\omega_0=\sqrt{\omega^2_R+\omega^2_I}, \label{omega0}
\end{eqnarray}
where the frequency of the harmonic oscillator becomes
$\omega_0=\omega_R$ for the case of long-lived quasinormal modes
$\omega_I\rightarrow0$ and becomes $\omega_0=|\omega_I|$ for the
most interesting case of highly excited quasinormal modes
$\omega_I\gg\omega_R$. We are interested in highly excited black
holes, i.e. $n$ is large, hence the proper frequency should be
$\omega_0=|\omega_I|$. In this framework, we consider the transition
$n\rightarrow n-1$ for a near-extremal SdS black hole. Thus
according to Eq. (\ref{QNM1}) and (\ref{omega0}), the absorbed
energy is
\begin{eqnarray}
\delta M&=&\hbar[(\omega_0)_n-(\omega_0)_{n-1}]\nonumber\\
&=&\hbar[(\omega_I)_{n-1}-(\omega_I)_n]\nonumber\\
&=&\hbar\kappa_+,\label{deltam0}
\end{eqnarray}
i.e. the transition frequency is
\begin{eqnarray}
\omega=[(\omega_I)_{n-1}-(\omega_I)_n]=\kappa_+.
\end{eqnarray}
Therefore, the adiabatic invariant (\ref{action}) can be rewritten
as
\begin{eqnarray}
I=\int\frac{d E}{\omega}=\int\frac{d
M}{\omega}=\frac{M}{\kappa_+}+c,
\end{eqnarray}
where $c$ is a constant. Bohr-Sommerfeld quantization then implies
that the mass spectrum is equally spaced, namely
\begin{eqnarray}
M=n\hbar\kappa_+.
\end{eqnarray}
We set $r_c-r_+=\Delta r$. By using Eqs. (\ref{m}) and
(\ref{deltam0}) we get
\begin{eqnarray}
\delta M=\frac{3r_+\Delta r\delta
r_+}{2L^2_{ds}}=\hbar\kappa_+,\label{deltam1}
\end{eqnarray}
On the other hand, the black hole horizon area is given by
\begin{eqnarray}
A_+=4\pi r^2_+,
\end{eqnarray}
Implementing the variation of the black hole horizon and using Eq.
(\ref{deltam1}), we have
\begin{eqnarray}
\delta A_+=8\pi r_+\delta
r_+=8\pi\frac{2L^2_{ds}\hbar\kappa_+}{\Delta r}.\label{deltaa}
\end{eqnarray}
Substituting Eq. (\ref{kappa}) into Eq. (\ref{deltaa}), one can
obtain
\begin{eqnarray}
\delta A_+=8\pi\hbar.
\end{eqnarray}
Hence, the area spectrum of near-extremal SdS black hole has the
following form:
\begin{eqnarray}
A_n=8\pi n\hbar.
\end{eqnarray}
According to the definition of the Benkenstein-Hawking entropy, we
have
\begin{eqnarray}
S=\frac{A_n}{4\hbar}=2\pi n.
\end{eqnarray}

Thus, in the highly damping limit, we have derived the area and
entropy spectrum of the near-extremal SdS black hole. In particular,
we find that the spectra is equally spaced. In addition, for the
eikonal limit ($l\gg1$), large-$l$ modes will dominate the black
hole's response to perturbations. Using Eq. (\ref{QNM2}) and the new
proposal, we have $\omega_0=\omega_R$ and the transition frequency
$\omega=(\omega_R)_l-(\omega_R)_{l-1}=\kappa_+$ for $l\gg n$. In
this framework, we could also obtain the same area spectrum.

\section{Conclusion}

In this paper, we have succeeded in utilizing a new physical
interpretation of quasinormal modes which was proposed by Maggiore
\cite{maggiore} to the near-extremal Schwarzschild de Sitter black
hole. According to Maggiore's proposal, we consider
$\omega=|\omega_I|_n-|\omega_I|_{n-1}$ to be the frequency of the
adiabatic invariant of a black hole that appears in the context of
Kunstatter's method, since we are working with the highly damped
quasinormal modes. In this framework, we have obtained the area and
entropy spectrum of event horizon which is different from Setare's
result \cite{setare2}. In addition, it is noteworthy that the
derived area spectrum of near-extremal SdS black holes in the four
dimensional spacetime is equally spaced which coincides with
Bohr-Sommerfeld quantization condition. On the other hand, the
derived quantum area is $\Delta A=8\pi\hbar$. As one can see in Ref.
\cite{maggiore,vagenas}, $\Delta A=8\pi\hbar$ for Schwarzschild and
Kerr black holes. Therefore, consistent with claim of the paper
\cite{andersson}, the quantum area $\Delta A$ may be universal for
all black holes. Furthermore, it would be of interest to generalize
our work to other black holes. Our result will also be supported by
observable effects in the future.

\begin{acknowledgments}
This work is supported by NSF (10703001), SRFDP (20070141034) of
P.R. China.
\end{acknowledgments}

\end{document}